\documentclass[journal=jacsat,manuscript=article]{achemso}
\usepackage[version=3]{mhchem} 
\usepackage{multirow}

\usepackage[dvipsnames]{xcolor}
\author{Diana Shakirova}
\affiliation{Institute of Physics, University of Graz, and NAWI Graz, Universit\"{a}tsplatz 5, Graz 8010, Austria}
\author{Adri\`a Can\'os Valero}
\affiliation{Institute of Physics, University of Graz, and NAWI Graz, Universit\"{a}tsplatz 5, Graz 8010, Austria}
\alsoaffiliation{Riga Technical University, Institute of Telecommunications, Riga, 1048, Latvia}
\author{Daniil Riabov}
\affiliation{Laboratory of Bionanophotonic Systems, Institute of Bioengineering, École Polytechnique Fédérale de Lausanne (EPFL), Lausanne 1015, Switzerland}
\author{Hatice Altug}
\affiliation{Laboratory of Bionanophotonic Systems, Institute of Bioengineering, École Polytechnique Fédérale de Lausanne (EPFL), Lausanne 1015, Switzerland}
\author{Andrey Bogdanov}
\email{a.bogdanov@hrbeu.edu.cn}
\affiliation{Qingdao Innovation and Development Center of Harbin Engineering University, Qingdao, 266500, China}
\alsoaffiliation{School of Physics and Engineering, ITMO University, 191002, St. Petersburg, Russia}
\author{Thomas Weiss}
\email{thomas.weiss@uni-graz.at}
\affiliation{Institute of Physics, University of Graz, and NAWI Graz, Universit\"{a}tsplatz 5, Graz 8010, Austria}
\title{Molecular chiral response enhanced by crosstalking quasi-bound states in the continuum}

\abbreviations{IR,NMR,UV}

\begin{document}


\begin{abstract}
Identifying the handedness of chiral molecules is of fundamental importance in chemistry, biology, pharmacy, and medicine. Nanophotonic structures allow us to control light at the nanoscale and offer powerful tools for chiral sensing, enabling the detection of small analyte volumes and low molecular concentrations by harnessing optical resonances. Most existing strategies rely on intuitive concepts such as strong local field enhancement or large local optical chirality, often achieved by engineering electric and magnetic Mie resonances in dielectric or plasmonic nanostructures. Recent insights, however, reveal that the chiroptical response of resonant systems is governed not only by local field effects, but also by less obvious mechanisms such as modal crosstalk. In this work, we present a dielectric metasurface engineered to amplify the modal crosstalk by supporting two nearly degenerate, high-quality-factor resonant states known as quasi-bound states in the continuum. Our theoretical and numerical analysis predicts a pronounced differential transmittance that exceeds the detection threshold of standard spectrometers. In particular, the differential transmittance reaches up to  $10^{-2}$ for the Pasteur parameter $\kappa = 1\cdot10^{-4}$. These findings advance the capabilities of nanophotonic sensors for chiral detection, paving the way toward ultrasensitive identification of molecular handedness in increasingly smaller volumes and concentrations at the experimentally visible level.
\begin{figure}
\includegraphics{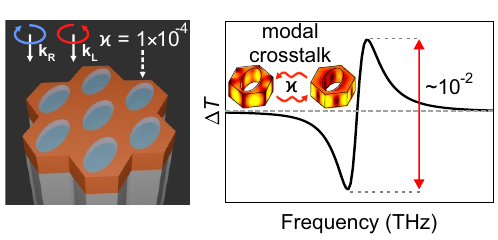}
\label{fig:concept}
\end{figure}
\end{abstract}

\section{Introduction}
Chirality is known as the non-identity of an object to its mirror image. This fundamental property is inherent in many organic molecules, chemical compounds, and drugs\cite{thall1996drug, woody_origin_1971, hembury_chirality-sensing_2008}. Although chirality seems to be a purely geometrical characteristic, it crucially affects living organisms. The same molecules of opposite handedness (enantiomers) can act as a drug or as a toxin, respectively \cite{thall1996drug}. Therefore, the detection of molecular handedness is a cornerstone problem of biology, chemistry, medicine, pharmacy, and food science \cite{ding_chiral_1997, nguyen_chiral_2006, dorazio_chiral_2017, alvarez-rivera_chiral_2020, wang_electrochemiluminescence_2023, jeong_breakdown_2020}.

One of the most well-known approaches of enantiomeric discrimination is to measure circular dichroism (CD), the difference in absorbance of left- and right-handed circularly polarized (LCP and RCP) light by chiral molecules. Unfortunately,  CD signals for natural chiral matter are very weak in the visible and infrared spectral ranges, so that they are below the detection limit for conventional CD spectroscopy for many practical applications, where only a limited number of molecules is available. The detection limit is typically around $\approx10^{-5}$ in terms of absolute transmittance or absorbance difference \cite{nesterov2016role}. This is where nanophotonics comes into play. The use of resonances of plasmonic and dielectric nanostructures has been shown to significantly enhance chiral light-matter interaction and increase CD \cite{govorov_theory_2010, govorov_plasmon-induced_2011, zhang_giant_2013, hendry_ultrasensitive_2010, hendry_chiral_2012, tullius_superchiral_2015, nesterov2016role, garcia-guirado_enantiomer-selective_2018, mohammadi_nanophotonic_2018, mohammadi_accessible_2019, chen_integrated_2020, mohammadi_dual_2021,  garcia-guirado_enhanced_2020, yoo2019metamaterials, li_dielectric_2023}. These resonances are associated with the resonant states (or quasinormal modes) of the optical systems that can trap the light efficiently at certain incident frequencies \cite{weiss_how_2018,lalanne_light_2018}. The quasinormal modes are characterized by complex eigenfrequencies, whose imaginary parts are associated with the total losses as a sum of radiative and absorption ones. \cite{lalanne_light_2018,both2022nanophotonic}.

One way to characterize chiral light-matter interaction involves evaluation of the so-called local optical chirality $C$ \cite{tang_optical_2010}. This quantity is proportional to the imaginary part of the scalar product between the electric and magnetic field, namely $C \sim \text{Im}(\textbf{E}\cdot\textbf{H}^*)$\cite{tang_optical_2010}, which is related to the absorbance difference between LCP and RCP light interacting with molecules at a certain position in space \cite{tang_optical_2010, hendry_ultrasensitive_2010, hendry_chiral_2012, schaferling_reducing_2016, mohammadi_dual_2021, garcia-guirado_enantiomer-selective_2018}. Thus, it characterizes the degree of local field chirality, which motivated many studies of the chiral fields generated by \textit{plasmonic} nanostructures in the first place \cite{govorov_theory_2010, govorov_plasmon-induced_2011, zhang_giant_2013, hendry_ultrasensitive_2010, hendry_chiral_2012,  tullius_superchiral_2015, nesterov2016role, garcia-guirado_enantiomer-selective_2018,schaferling_tailoring_2012, schaferling_helical_2014,  schaferling_reducing_2016}. The resonant states of the latter guarantee strong field enhancements within small volumes in the optical spectral range and may be used to achieve a locally large so-called optical chirality $C$. Both single particles \cite{govorov_theory_2010, govorov_plasmon-induced_2011, zhang_giant_2013} and periodic arrays \cite{hendry_ultrasensitive_2010, hendry_chiral_2012,  nesterov2016role, tullius_superchiral_2015,garcia-guirado_enantiomer-selective_2018} have demonstrated amplified chiral response from a sample placed in their vicinity. In particular, significant CD enhancement of the order of $10^{3}$ has also been reported in \cite{nesterov2016role} for the case of a uniform chiral analyte placed at the edges of plasmonic nanorods. However, even this enhancement does not beat the detection threshold of conventional spectrometers when considering a restricted analyte volume.

In recent years, \textit{dielectric} nanophotonics has emerged as a promising alternative to plasmonics \cite{yoo2019metamaterials, mohammadi_nanophotonic_2018, mohammadi_accessible_2019, garcia-guirado_enhanced_2020, chen_integrated_2020, mohammadi_dual_2021, droulias_absolute_2020, guan_ultrasensitive_2025, almousa_employing_2024}. Both theoretical \cite{mohammadi_nanophotonic_2018, mohammadi_accessible_2019, mohammadi_dual_2021, chen_integrated_2020} and experimental \cite{garcia-guirado_enhanced_2020} works have shown the effect of CD enhancement on the order of $10-10^{2}$ by using all-dielectric or metal-dielectric platforms. The first reason of such efficiency is the ability to harness the magnetic resonances in dielectric structures, which in proper combination with the electric ones can significantly increase the local optical chirality \cite{mohammadi_nanophotonic_2018, mohammadi_accessible_2019, garcia-guirado_enhanced_2020, mohammadi_dual_2021, chen_integrated_2020, li_dielectric_2023, li_enhanced_2024, almousa_employing_2024}. The second reason is long temporal confinement of light in dielectric materials. Quantitatively, the duration of light confinement by a resonant state can be characterized by the quality factor $Q$, which is minus the ratio of the real and twice the imaginary parts of its eigenfrequency $\omega_n$, $Q = \left|\text{Re}(\omega_n)/2\text{Im}(\omega_n)\right|$. Resonant states with significantly higher $Q$ factors are accessible in dielectric nanoresonators due to smaller intrinsic material losses compared to plasmonic platforms. 

Special cases of resonant states possessing infinite $Q$ factors are bound states in the continuum (BICs). Originally proposed in quantum mechanics \cite{von_neumann_uber_1993}, the concept of BICs has been transferred to optics recently \cite{koshelev_meta-optics_2019, koshelev2020subwavelength}. However, the advantage of infinitely high $Q$ factors of these resonant states cannot be used since BICs are uncoupled from free space plane waves. One way to overcome this obstacle is to involve deformation or structural imperfections, thus introducing radiative losses in a resonator and allowing for the plane waves outcoupling \cite{koshelev_asymmetric_2018}. In combination with intrinsic material losses, radiative losses limit the $Q$ factors of the perturbed modes, which are referred to as so-called quasi-bound states in the continuum (quasi-BICs) – resonant states possessing finite, but still significantly high $Q$ factors. Quasi-BICs have attracted a lot of attention in the nanophotonics community over the last decades and have demonstrated an enhanced performance of devices in lasing \cite{kodigala_lasing_2017, ha_directional_2018}, active photonics \cite{han_all-dielectric_nodate}, biophotonics \cite{romano_label-free_2018, romano_surface-enhanced_2018}, and polaritonics \cite{koshelev_strong_2018}. In the chiral sensing community, this concept has also been used for notable amplification of chiroptical response \cite{chen_integrated_2020, guan_ultrasensitive_2025, wang_optical_2024}. In Ref.~\citenum{guan_ultrasensitive_2025}, a huge enhancement  of $10^5$ order is stated for a metasurface that supports nearly degenerate quasi-BICs. Extremely high quality factors of resonant states $Q \approx 10^6$ are obtained for the lossless system, which is, however, challenging to realize experimentally \cite{chen_observation_2022}. In addition, BICs have also been used to design metasurfaces exhibiting high local optical chirality \cite{gorkunov_metasurfaces_2020, zhang_chiral_2022, chen_observation_2023}.

In spite of the discussed huge progress in the field of chiral sensing, only recently, a general electromagnetic theory of chiral light-matter interaction in arbitrary resonators has been developed \cite{both2022nanophotonic}. The idea behind it is the ability to reconstruct the optical response (i.e., transmission, reflection and absorption) from a system via its resonant states \cite{lalanne_light_2018, lobanov_resonant-state_2018, 
weiss_how_2018, zhang_quasinormal_2020}. Moreover, one can predict the change in the optical response with a small perturbation introduced in the system, which is, in our case, given by the chiral analyte of interest. In the framework of this theory, this change can be disentangled into a sum of four contributions, namely, resonance shift, changes in the excitation and emission of resonances, and \textit{modal crosstalk}. The resonance shift describes how strongly a particular resonance of an unperturbed structure spectrally shifts after inserting a chiral analyte. This contribution is present only in geometrically chiral structures \cite{both2022nanophotonic}, and has already been well studied \cite{hendry_ultrasensitive_2010, hendry_chiral_2012, tullius_superchiral_2015}. The second and third contributions are responsible for the change in optical response intensity between unperturbed and perturbed resonator. This effect has also been realized in many works \cite{nesterov2016role, schaferling_reducing_2016}. The last mechanism, the modal crosstalk, implies any alteration of the optical spectrum caused by the interplay of spectrally close resonant states due to the presence of a chiral medium. While the first three processes have been widely realized, albeit without explicit recognition in most studies, the modal crosstalk for chiral sensing is rather unexploited. Although some works have combined electric and magnetic modes,\cite{mohammadi_nanophotonic_2018, garcia-guirado_enhanced_2020, guan_ultrasensitive_2025, almousa_employing_2024} so far, there was no strict recipe to maximize this contribution, which, in particular, allows us to benefit from a bisignate chiroptical response for the real-dominated Pasteur parameter, as shown further. 

In this work, we investigate possibilities to maximize the effect of the modal crosstalk. Notably, we uncover that the latter can be significantly boosted by leveraging the interaction between two well-designed high $Q$-factor eigenmodes. In particular, the eigenmodes must have opposite parity under inversion, and their resonant frequencies must be nearly degenerate. In contrast with Ref.~\citenum{guan_ultrasensitive_2025}, we operate in the weak-coupling regime of the eigenmodes, which allows us to apply the electromagnetic theory of chiral light-matter interaction \cite{both2022nanophotonic}. Here, we demonstrate that a proper combination of two quasi-BICs guarantees much vaster enhancement of CD and provide a robust theoretical explanation of the effect.

We start our exploration with the analysis of the expression for the modal crosstalk contribution derived in Ref.~\citenum{both2022nanophotonic} and determine conditions for its maximization. Further, we design a metasurface corresponding to the defined requirements and accounting for experimental restrictions, and confirm by both full-wave simulations and the modal theory the predicted CD enhancement. Besides physical insights explaining the modal crosstalk, we obtain a differential transmittance on the order of $10^{-2}$ for $\kappa = 1\cdot10^{-4}$, which provides a promising route to beat the detection limit for small analyte volumes and low concentrations of chiral molecules. We also compare our metasurface to existing plasmonic\cite{both2022nanophotonic} and dielectric platforms\cite{mohammadi_nanophotonic_2018, garcia-guirado_enhanced_2020}, proving the benefits of our design. 

\section{Results and discussion}

We start our discussion by formulating a theory of CD induced by a resonant structure. To efficiently characterize the interaction of incident light with a resonator, the optical scattering matrix $S$ is widely used \cite{lobanov_resonant-state_2018, weiss_how_2018, zhang_quasinormal_2020, alpeggiani_quasinormal-mode_2017}. In the planar systems considered here, each element of the scattering matrix, $S_{\mathbf{MN}}$, represents the amplitude for light to scatter from an incoming channel $\mathbf{N}$ to an outgoing channel $\mathbf{M}$, through either transmission or reflection. Here, we adopt the conventions from Ref.~\citenum{both2022nanophotonic} and use the indices $\mathbf{N}$ and $\mathbf{M}$ as sets of quantum numbers that characterize a particular scattering channel. Each channel is defined by two parameters: the polarization of the light — left- or right-handed circularly polarized (L or R) — and the region of propagation — top (t) or bottom (b), denoting the superstrate or the substrate. For example, the matrix element $|S_{\text{Rb}, \text{Lt}}|^2$, which is also written as $|t_{\text{RL}}|^2$ or $T_{\text{RL}}$, describes the transmittance of left-handed circularly polarized (LCP) light incident from the top into right-handed circularly polarized (RCP) light transmitted to the bottom. In periodic systems, $\mathbf{N}$ and $\mathbf{M}$ may also include diffraction orders, since scattering can occur into multiple diffraction channels. However, only those channels corresponding to propagating in the free space waves are typically considered. For arrays with subwavelength periodicity, only the zeroth diffraction order satisfies this condition. As our study is limited to this regime, we omit diffraction orders from the notation of the quantum numbers $\mathbf{N}$ and $\mathbf{M}$.

If an arbitrary perturbation, which is a chiral analyte in our case, is inserted into a resonator, the scattering matrix changes from $S$ to $S + \delta S$, where $\delta S$ corresponds to the change upon the perturbation. The latter can be calculated with a perturbation theory and expressed as a sum of the four terms associated with resonant effects: $\delta S = \delta S^{\text{ex}} + \delta S^{\text{em}} + \delta S^{\text{shift}} + \delta S^{\text{cross}}$ \cite{both2022nanophotonic}. These contributions are the change in excitation $\delta S^{\text{ex}}$ and emission $\delta S^{\text{em}}$, defined by the coupling of modes with incoming and outgoing fields, respectively; the resonance shift $\delta S^{\text{shift}}$, containing an overlap integral of the modes with themselves; and the modal crosstalk $\delta S^{\text{cross}}$, which takes into account the interaction between different modes.

\sloppy
In general, CD is defined as the difference in absorbance $\Delta A$ for LCP and RCP light \cite{schaferling_helical_2014, nesterov2016role,schaferling_chiral_2017, schaferling_reducing_2016, both2022nanophotonic}. The differential absorbance $\Delta A$ is related to the differential transmittance $\Delta T$ and the differential reflectance $\Delta R$, which are differences in transmittance and reflectance for LCP and RCP light, respectively, as CD~$=\Delta A = -\Delta T-\Delta R$. In special cases, such as a homogeneous medium or periodic arrays possessing symmetry $C_3$ and higher \cite{bai_determination_2012, schaferling_chiral_2017}, $\Delta R$ goes to zero. Therefore, one can associate CD with $\Delta T$ straightforwardly, which is also more feasible for experimental measurements \cite{mohammadi_accessible_2019, mohammadi_nanophotonic_2018, garcia-guirado_enantiomer-selective_2018, garcia-guirado_enhanced_2020}. In this work, we propose a structure that demonstrates rather low differential reflectance, which allows us to use $\Delta T$ as a figure of merit (FOM). Furthermore, one should take into account circular polarization conversion (CPC) that appears for LCP and RCP light in structures possessing lower then $C_3$ symmetries\cite{schaferling_chiral_2017, bai_determination_2012}. Caused by elliptical birefringence, CPC can be assumed erroneously to be CD \cite{schaferling_chiral_2017}. Therefore, we choose  $\Delta T = T_\text{LL} -T_\text{RR}$ as a FOM to be maximized, with the assumption of CPC to be negligibly small, i.e.,  $T_\text{RL}, T_\text{LR} \ll T_\text{LL}, T_\text{RR}$, and  $T_\text{RL} - T_\text{LR} \ll T_\text{LL} - T_\text{RR}$. Thus, the differential transmittance can be written as:
\begin{equation}
    \Delta T = |S_\text{Lb,Lt}|^2 - |S_\text{Rb,Rt}|^2 = |S_\text{Lb,Lt}^{(0)} + \delta S_\text{Lb,Lt}|^2 - |S_\text{Rb,Rt}^{(0)} + \delta S_\text{Rb,Rt}|^2,
    \label{eq: DT_pert}
\end{equation}
with $S^{(0)}$ and $\delta S$ being the unperturbed transmission amplitude and its perturbation correction, respectively.

We dedicate this work to the exploration of the perturbation in transmission amplitude dominated by modal crosstalk, namely $\delta S \approx \delta S^{\text{cross}}$, which is found as \cite{both2022nanophotonic}

\begin{equation}
    \delta S^{\text{cross}}_{{\textbf{MN}}} = \sum_{n\neq n'}\frac{a_{n,\textbf{M}}b_{n',\textbf{N}}\int_{V_c}i\omega c\kappa\left(\textbf{E}_n\cdot\textbf{H}_{n'}+\textbf{H}_n\cdot\textbf{E}_{n'}\right)\text{dV}}{(\omega - \omega_n)(\omega - \omega_{n'})},
\label{eq: deltaS_cross}
\end{equation}
where $a_{n,\textbf{M}}$ and $b_{n',\textbf{N}}$ are emission and excitation coefficients of the modes $n$ and $n'$, respectively (see Ref.~\citenum{both2022nanophotonic} for definition), $\omega$ and $\omega_n$ are, respectively, the angular frequency of the incident light and the eigenfrequency of the resonant state with index $n$ (do not mix up with refractive index). Furthermore, $c$ is the speed of light in vacuum, and $\textbf{E}_n$ and $\textbf{H}_n$ are electric and magnetic resonant field distributions. Finally, $\kappa$ is the Pasteur parameter describing the chirality strength of an isotropic and homogeneous molecular solution \cite{nesterov2016role, schaferling_reducing_2016, schaferling_chiral_2017}. An opposite sign of $\kappa$ corresponds to opposite handedness of the molecular solution.

We now establish the prerequisites for $\Delta T$ to be maximized by modal crosstalk $\delta S^{\text{cross}}$. First, we have assumed a resonator supporting two resonant states with indices $n$ and $n'$ that are spectrally close to each other, i.e., the linewidths of the corresponding resonances overlap. It is evident from Eq.~(\ref{eq: deltaS_cross}) that degeneracy of the latter would guarantee a sharp drop of the denominator: $\text{Re}(\omega_n) = \text{Re}(\omega_{n'})$ and $\omega - \text{Re}(\omega_n) \approx 0$ in the vicinity of the resonance. Therefore, the only non-zero contribution to the denominator is the product of $\text{Im}(\omega_n)\text{Im}(\omega_{n'})$, which can be significantly reduced with the use of high-$Q$ modes. Second, the choice of orthogonally polarized resonant states enhances the integrand in Eq.~(\ref{eq: deltaS_cross}), namely, the scalar products of electric and magnetic modal fields $\textbf{E}_n\cdot\textbf{H}_{n'}$ and $\textbf{E}_{n'}\cdot\textbf{H}_n$ are maximized. Thus, spectral degeneracy of orthogonally polarized high-$Q$ modes boost $\delta S^{\text{cross}}$. To efficiently exploit the modes, one also has to maximize the frequency-dependent emission and excitation coefficients $a_{n,\textbf{M}}(\omega)$ and $b_{n',\textbf{N}}(\omega)$, which represent the coupling of the modal fields to the outgoing and incoming waves, respectively \cite{both2022nanophotonic}. Therefore, $a_{n,\textbf{M}}$ and $b_{n',\textbf{N}}$ can be enhanced by a proper choice of incoming and outgoing waves polarization, which increases the scalar product of the modal fields with the latter. In the next step, we show how the established prerequisites can be realized in a realistic metasurface design.

\begin{figure}
\includegraphics[width=0.8\linewidth]{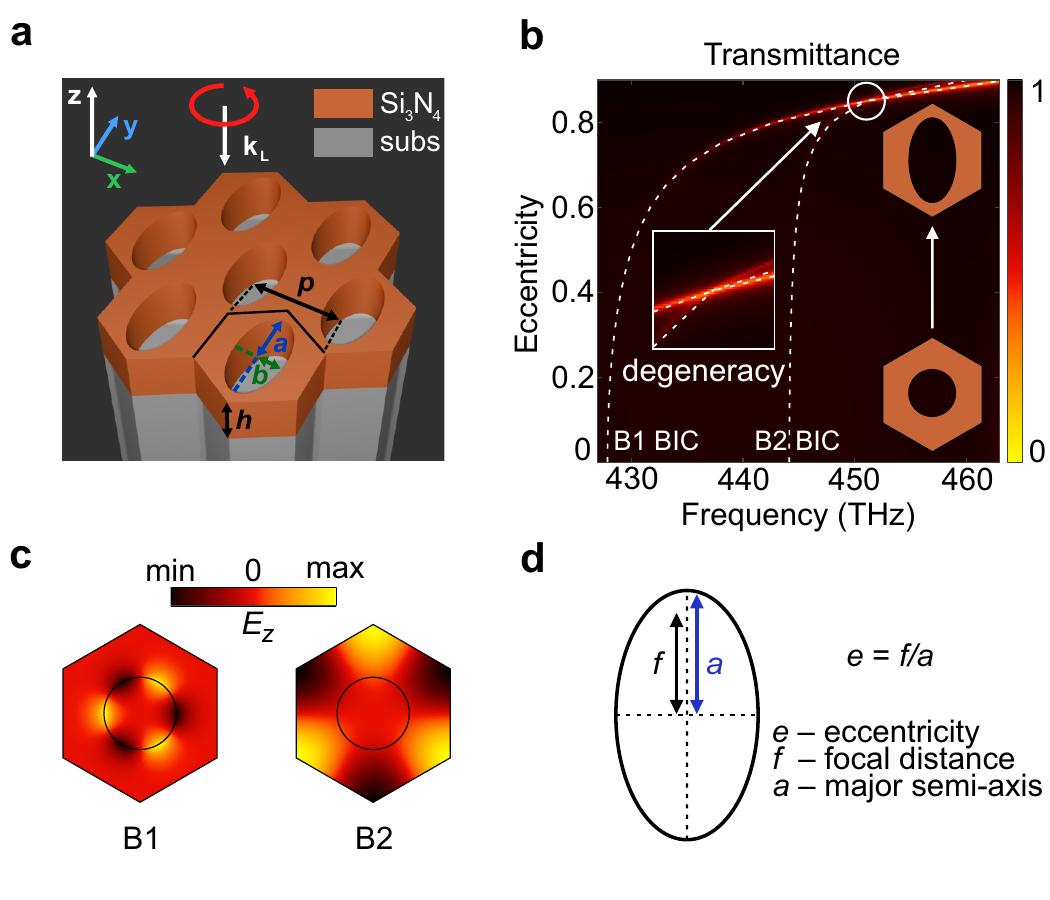}
\caption{(a) Metasurface design as a periodic array of silicon nitride \ce{Si3N4} slab with triangular unit cells and refractive index $n_{\ce{Si3N4}} = 1.9464 + 0.0012i$ on a model substrate with $n_{\text{subs}} = 1.4$. The unit cell with period $p = 470$~nm and height $h = 230$~nm contains elliptical void characterized by $a = 208.5$~nm and $b = 110$~nm as large and small semi-axes, respectively. \mbox{(b) Transmittance} map for the metasurface illuminated by left-handed circularly polarized light. Transmittance is calculated by full-wave simulations as a function of frequency and eccentricity, which is defined in panel (d). Dashed white lines represent the dispersion of the considered quasi-bound states in the continuum B1 and B2 calculated by eigenvalue solver. For $e \approx 0.85$, a degeneracy in real parts of quasi-bound states in the continuum eigenfrequencies is observed. (c) Distribution of $E_z$ as the $z$ component of the electric field in the undeformed unit cell for B1 and B2. (d) Definition of the eccentricity: The initially circle-shaped cross section of the unit cell void is  deformed by stretching along the y axis. The eccentricity is as defined $e = f/a$ with $f$ and $a$ being focal distance and the major semi-axis of the ellipse.}
\label{fig:1}
\end{figure}

We propose a metasurface with a triangular unit cell containing a circular void. The metasurface is made of silicon nitride (\ce{Si3N4}) and placed on a transparent model substrate. 
We set the upper half-space to be water with $n_{\ce{H2O}} = 1.33$, which is not shown in Fig.~\ref{fig:1}(a) for illustrative reasons. This configuration resembles a racemic mixture of chiral molecules so that any spectral shift of the bare metasurface modes only originates in the deviation from this racemic mixture.

The system exhibits $C_{6v}$ symmetry at normal incidence and supports symmetry-protected BICs \cite{overvig_selection_2020}. In particular, the metasurface demonstrates two BICs at 428~THz ($\approx$ 701~nm) and 444~THz ($\approx$ 676~nm) that are associated to the B1 and B2 irreducible representations, respectively \cite{overvig_selection_2020} [Fig.~\ref{fig:1}(c)]. The modes transforming under these representations possess opposite parity under inversion and guarantee electric fields of the modes to be orthogonal to each other, aligning with our aim. Theoretically, the BICs exhibit an infinite $Q$ factor \cite{von_neumann_uber_1993, koshelev_meta-optics_2019, koshelev2020subwavelength}, which also makes them perfect candidates for $\Delta T$ boosting. Further, we break the symmetry from $C_{6v}$ to $C_{2v}$ by stretching the circular void to an elliptical one as depicted in Fig.~\ref{fig:1}(b).  With such deformation, we  perturb the resonant states and couple them to free space, transforming BICs to quasi-BICs \cite{koshelev_meta-optics_2019, koshelev2020subwavelength, kang_merging_2022}. The deformation strength can be characterized by the hole eccentricity, introduced in Fig.~\ref{fig:1}(d). Additionally, this deformation allows us to tune the dispersion of quasi-BICs. Fig.~\ref{fig:1}(b) shows a color map for the transmittance as a function of frequency and eccentricity. White dashed lines represent the dispersion of B1 and B2 BICs. Illumination under normal incidence with LCP light allows us to excite both orthogonally linearly polarized modes simultaneously.

We keep notations B1 and B2 throughout this manuscript to refer to the corresponding quasi-BICs, since the $C_{2v}$ group still contains the B1 and B2 irreducible representations. Moreover, as a result, the modes do not couple with each other with the chosen perturbation. The degeneracy in real parts of their eigenfrequencies is observed for the eccentricity $e\approx0.85$. The metasurface geometrical parameters corresponding to the mode degeneracy are depicted in Fig.~\ref{fig:1}(a) and specified in the caption. To calculate the transmittance dependence on the eccentricity, only the major semi-axis $a$ is varied, while all other parameters are fixed.

\begin{figure}
\includegraphics[width=0.8\linewidth]{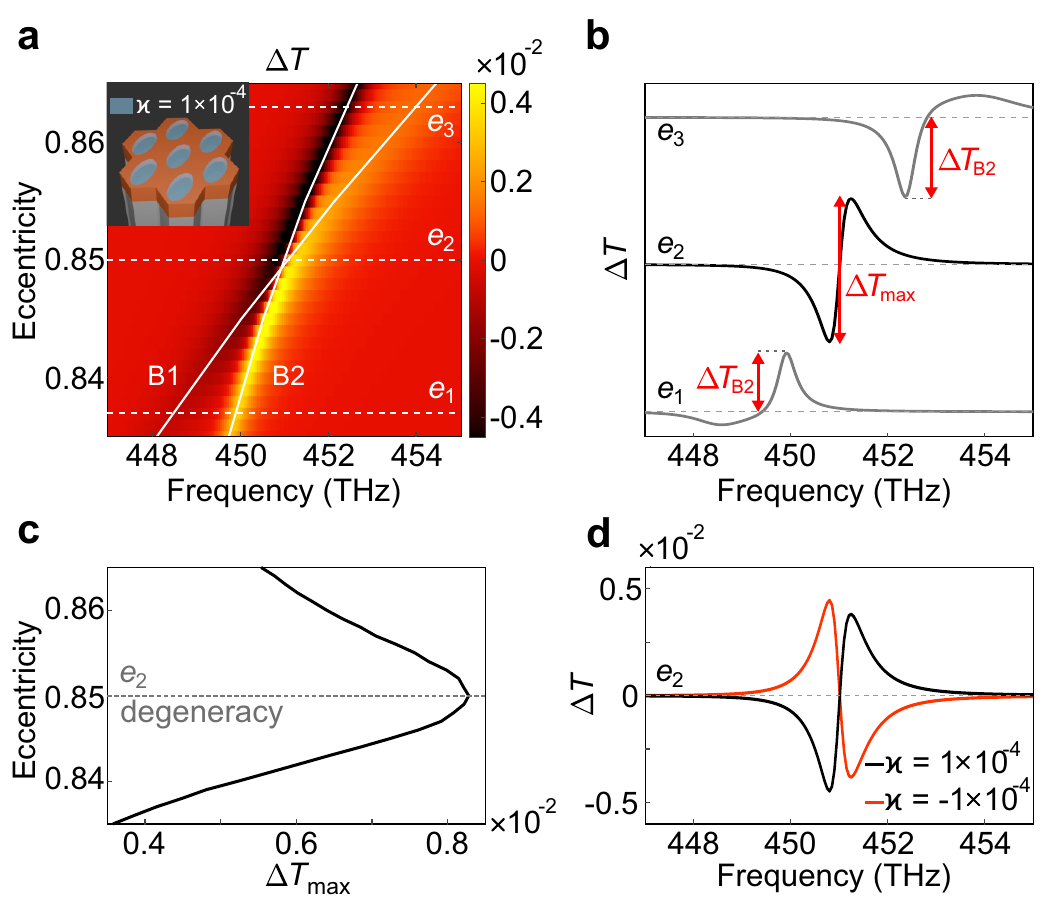}
\caption{$\Delta T$ response of a chiral analyte placed on the metasurface. (a) Colormap for $\Delta T$ as function of incident frequency and eccentricity. Solid white lines show the dispersion of B1 and B2 modes. Dashed lines highlight the eccentricity values to be discussed, and $e_2\approx0.85$ in particular corresponds to the degeneracy. The inset sketches the metasurface with the chiral analyte placed into the voids. The Pasteur parameter here is $\kappa = 1\times10^{-4}$. (b) Examples of $\Delta T$ for $e_1\approx0.84$, $e_2$, and $e_3\approx0.86$. It is seen that the amplitude of the bisignate response $\Delta T_{\text{max}}$ observed at the degeneracy at $e_2$ significantly exceeds chiral responses associated with single quasi-BICs located further from the degeneracy at $e_1$ and $e_3$. (c) Amplitude $\Delta T_{\text{max}}$ as a function of eccentricity. The highest value corresponds to the degeneracy at $e_2$. (d) $\Delta T$ calculated for two Pasteur parameter values with opposite signs. Sign switch of the spectrum guarantees that $\Delta T$ is provided by the chiral nature of analyte, proving the applicability of the system for handedness discrimination.}
\label{fig:2}
\end{figure}

Once the proper parameters of the metasurface are chosen, we add a chiral analyte in the system. Here, we introduce a chiral analyte with the Pasteur parameter $\kappa = 1\times10^{-4}$, which is the upper limit of realistic values used in literature \cite{garcia-guirado_enhanced_2020, both2022nanophotonic}. In the Supporting Information, we show that for the chiral analyte with $\text{Re}(\kappa) \gg \text{Im}(\kappa)$, which is a typical case in the visible frequency range \cite{nesterov2016role, both2022nanophotonic, garcia-guirado_enhanced_2020, mohammadi_nanophotonic_2018}, the imaginary part of the Pasteur parameter plays a crucial role only for the bare chiral analyte and the corresponding enhancement factor, while negligibly affects the absolute chiroptical response induced by the metasurface. Therefore, we fix the real-valued Pasteur parameter, which allows us to use a sparser mesh to reduce the computational cost of the calculations. The additional degree of freedom is the amount of chiral solution to be used. On the one hand, enough space should be occupied by the chiral medium to make use of all near fields provided by the metasurface, maximizing the numerator in Eq.~(\ref{eq: deltaS_cross}). On the other hand, if there is too much chiral analyte, the chiroptical response becomes dominated not by the resonator enhancement but by the extra amount of molecular solution instead. To account for these aspects, the chiral sample is inserted inside elliptical voids only, completely filling the later [inset in Fig.~\ref{fig:2}(a)]. A detailed discussion of the optimal amount of the solution to be used is provided in the Supporting Information (Fig.~S4).

The chiroptical response of the system in the vicinity of the degeneracy is plotted  as a function of frequency and eccentricity in Fig.~\ref{fig:2}(a). Here, white solid lines show the dispersion of the quasi-BICs, and the color represents $\Delta T$. To prove the advantage of the degenerate configuration, we investigate $\Delta T$ at eccentricities below, above and in close proximity to the degeneracy. In Fig.~\ref{fig:2}(b), $\Delta T$ spectra corresponding to the eccentricity values $e_1\approx0.84$, $e_2\approx0.85$, and $e_3\approx0.86$ are shown. It is clearly seen that the case of degenerate resonant states allows us to make use of the bisignate signal shape. The latter is originally known as bisignate Cotton effect in supramolecular systems \cite{hembury_chirality-sensing_2008}, and partially discussed in \cite{droulias_chiral_2020, droulias_absolute_2020} by Droulias and Bougas. We label the peak to peak signal amplitude as $\Delta T_{\text{max}}$, which significantly exceeds  $|\Delta T_{\text{B2}}|$ provided by the single quasi-BIC B2 demonstrated at $e_1$ and $e_3$. One can also measure not only single peak amplitude at $e_1$ and $e_3$, but peak to peak value $\Delta T_{\text{max}}$. In Fig.~\ref{fig:2}(c), the dependence of $\Delta T_{\text{max}}$ on the eccentricity is shown. We flipped the axes to keep them consistent with Fig.~\ref{fig:2}(a), but the message is clear: The highest $\Delta T_{\text{max}}$ is observed for the degenerate quasi-BICs. To make sure that the chiroptical response of the system is defined by intrinsic chirality of the analyte, we calculate $\Delta T$ for opposite signs of the Pasteur parameter. Indeed, $\Delta T$ flips its sign, as shown in Fig.~\ref{fig:2}(d).

\begin{figure}
\includegraphics[width=0.8\linewidth]{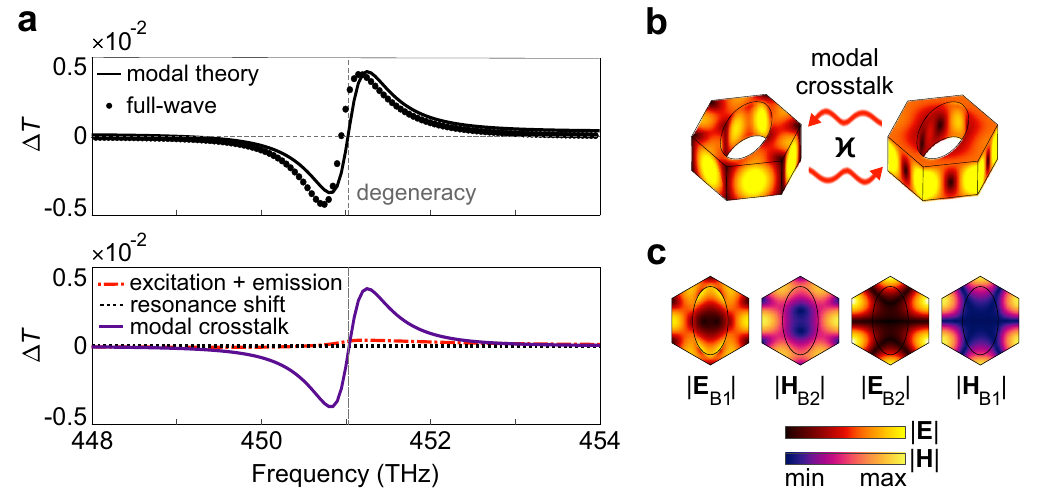}
\caption{Modal theory. (a) Modal theory calculations. The upper plot shows  $\Delta T$ predicted by modal theory (solid line) and full-wave calculation (dots). The results demonstrate good agreement. In the lower plot all mechanisms contributing to the $\Delta T$ are presented separately. As was expected, the dominant mechanism is the modal crosstalk defined in Eq.~(\ref{eq: deltaS_cross}). (b) Scheme of the modal crosstalk between the considered quasi-BICs, which is the main mechanism contributing to the enhancement of the chiral response. (c) Field distribution in the unit cell for B1 and B2 modes. Two colormaps are associated with absolute values of electric and magnetic fields. It is clearly seen that the electric and magnetic fields of different modes possess the same symmetry.}
\label{fig:3}
\end{figure}

Finally, we want to verify that the main mechanism of $\Delta T$ enhancement is the modal crosstalk. One can calculate all contributions $\delta S^{\text{ex}}$, $\delta S^{\text{em}}$, $\delta S^{\text{shift}}$, and $\delta S^{\text{cross}}$ straightforwardly from the modal fields according to the theory explained in Ref.~\citenum{both2022nanophotonic}. In Fig.~\ref{fig:3}(a), the results of the modal analysis are presented. The upper plot shows the comparison between the modal theory and numerical full-wave calculations. Notably, both approaches are in good agreement, despite the fact that only two modes B1 and B2 are used for the modal theory. The small deviation is caused by the background contributions of other modes located outside the considered spectral range. The lower panel depicts the contributions of individual terms. It is clearly seen that the modal crosstalk, schematically shown in Fig.~\ref{fig:3}(b), unambiguously dominates over the others. 

In addition, Fig.~\ref{fig:3}(c), shows the field distributions for the two modes at degeneracy. It can be seen that the electric field of B1 has the same intensity distribution pattern as B2. The same holds for the magnetic field of B1 and the electric field of B2. This fact ensures a maximized numerator in Eq.~(\ref{eq: deltaS_cross}).

In order to demonstrate the potential of our approach, we compare our results to several key works in chiral sensing \cite{both2022nanophotonic, mohammadi_nanophotonic_2018, garcia-guirado_enhanced_2020}. There exists a large body of literature claiming significant CD enhancements in complex designs, which have so far not been experimentally validated \cite{zhang_giant_2013, schaferling_reducing_2016, chen_integrated_2020, mohammadi_dual_2021, guan_ultrasensitive_2025}. To ensure a meaningful comparison, we have deliberately chosen to benchmark our results only against well-established designs and experimentally confirmed studies.

In Refs.~\citenum{mohammadi_nanophotonic_2018, garcia-guirado_enhanced_2020}, the authors presented metasurface designs supporting the combination of electric and magnetic dipoles. To obtain comparable values, we have transformed their results to the differential transmittance and estimated the corresponding enhancement to be
\begin{equation}
    \Delta T_\text{enh}=\frac{\Delta T}{\Delta T_\text{ch}},
    \label{eq: DT_enh}
\end{equation}
where $\Delta T_\text{ch}$ is the differential transmittance of the bare chiral analyte with the Pasteur parameter $\kappa = (1+0.01i)\times10^{-4}$, which is used in Refs.~\citenum{mohammadi_nanophotonic_2018, garcia-guirado_enhanced_2020}. For all designs, we take the highest enhancement obtained, including peak-to-peak amplitude for those that present bisignate spectra. We show that the triangular metasurface provides $10^2-10^3$ higher enhancement compared to these works.

We also compare our results with a prototypical periodic array of plasmonic nanorods, studied in Refs.~\citenum{both2022nanophotonic, nesterov2016role}. For $\kappa = (1+0.01i)\times 10^{-4}$, Both et al. obtain the differential absorbance enhancement $\Delta A_\text{enh} \approx 350$, while the same quantity achieved with the triangular metasurface is 1000-fold.  
\begin{table}
  \caption{Literature comparison. FOMs from key works in chiral sensing are transformed to $\Delta T_\text{enh}$ ($\Delta A_\text{enh}$) and compared to the triangular metasurface (TMS).}
  \label{tbl:example}
  \begin{tabular}{llll}
    \hline
    Paper  & FOM & Enhancement  \\
    \hline
    Mohammadi et al. (2018) & $\Delta T_\text{enh}$ & 4.2 \\
    García-Guirado et al. (2020)& $\Delta T_\text{enh}$ & 50\\
    Both et al. (2022) & $\Delta A_\text{enh}$ & 350 \\
    \hline
    \multirow{2}{*}{TMS} &$\Delta T_\text{enh}$&1950\\
    &$\Delta A_\text{enh}$&1000\\
    \hline
  \end{tabular}
  \label{tab: comparison}
\end{table}
\section{Conclusion}
We have presented a metasurface design for molecular chiral sensing based on nearly-de\-ge\-ne\-ra\-te quasi-BICs. The main mechanism driving the chiroptical response is the modal crosstalk. To maximize it, the general strategy has been theoretically established and confirmed by the modal theory and full-wave simulations. We have shown that in the special case of the degenerate, orthogonally-polarized quasi-BICs possessing opposite parity, the modal crosstalk mechanism becomes extremely dominant and significantly boosts the differential transmittance of the system. 

We have also compared our metasurface to experimentally-demonstrated \cite{garcia-guirado_enhanced_2020} and well-established designs \cite{mohammadi_nanophotonic_2018, both2022nanophotonic} in key works of chiral sensing, and unambiguously surpassed the previous results. It has been shown that our design provides significant enhancement with the FOM, which does not require any signal postprocessing. We believe that our work paves the way to an experimentally feasible approach in chiral sensing that will allow for reliable handedness discrimination of small analyte volumes and low concentrations.
\begin{acknowledgement}
Adri\`a Can\'os Valero acknowledges funding by the project No 1.1.1.9/LZP/1/24/101 : "Non-Hermitian physics of spatiotemporal photonic crystals of arbitrary shape (PROTOTYPE)".

The numerical simulations were partially supported  by the Russian Science Foundation (Project 23-72-10059).

Hatice Altug and Daniil Riabov would like to thank Swiss State Secretariat for Education, Research and Innovation (SERI) for financial support under the contract numbers of 22.00018 in connection to the projects from European Union’s Horizon Europe Research and Innovation Programme under agreements 101046424 (TwistedNano) and the European Union’s Horizon 2020 research and innovation programme under the Marie Skłodowska-Curie grant agreement No. 955623 (H2020MSCA-CONSENSE).
\end{acknowledgement}


\bibliography{achemso-demo}

\end{document}